\documentclass[10pt,conference,a4paper]{IEEEtran}
\usepackage{graphicx}

\begin{document}

\title{\huge{Fast and Precise 3D Computation of Capacitance of\\
Parallel Narrow Beam MEMS Structures}}

\author{
\large{N. Majumdar, S. Mukhopadhyay}\\
INO Section, Saha Institute of Nuclear Physics,
1/AF, Bidhannagar, Kolkata-700064, India\\
nayana.majumdar@saha.ac.in, supratik.mukhopadhyay@saha.ac.in\\}
\maketitle

\noindent{\large{\textbf{Abstract}}}\\
Efficient design and performance of electrically actuated MEMS devices 
necessitate accurate 
estimation of electrostatic forces on the MEMS structures. 
This in turn requires thorough study of the capacitance
of the structures and finally the charge density distribution on the various
surfaces of a device. In this work, nearly exact BEM
solutions have been provided in order to estimate these properties 
of a parallel narrow beam structure found in
MEMS devices. The effect of three-dimensionality, which is an important aspect
for these structures, and associated fringe fields have been studied in detail.
A reasonably large parameter
space has been covered in order to follow the variation of
capacitance with various geometric factors. The present results have been
compared with those obtained using empirical parametrized expressions 
keeping in
view the requirement of the speed of computation. The limitations of the
empirical expressions have been pointed out and possible approaches of their
improvement have been discussed.

\textbf{Keywords:} MEMS, narrow beam, comb drive, Boundary Element Method (BEM),
fringe capacitance.

\section{\large{\textbf{Introduction}}}
\noindent{The} capacitance in MEMS structures has always
been an important aspect to be studied
from the very beginning of this field of research. Under certain
circumstances, the presence of capacitance has turned out to be beneficial
while it has been considered as a challenge in some other. In
either case, estimation of this capacitance has drawn prime importance 
reasonably in the
design and subsequent use of these devices.
The sensitivity, instability and dynamics of a MEMS device depend crucially
upon an interplay of electrical and mechanical forces generated within the 
device. Since the electrostatic
force alters the dynamic properties of a MEMS structure and the electrostatic
charges redistribute as the structure deforms and thereby change 
the electrostatic
force distribution, a complex electromechanical coupling dictates the 
performance of the device. This may induce non-linearity in electrostatic force
ultimately leading to a pull-in stage. The presence of fringe 
field may further
complicate the situation. Thus an accurate device modeling calls for an
efficient electromechanical analysis of the structure which in turn depends
crucially on the precision of electrostatic analysis. The electrostatic 
analysis deals with precise estimation of observables like charge density 
distribution, total charge content etc. from which the important quantities
like capacitance, electrostatic force can be estimated.
While the capacitance of a conductor
represents its overall charge content and, thus, determines several important
properties like electrostatic load distribution on a given structure or 
induced current, 
the charge distribution on the conductor finally becomes the most important
observable. 

The approaches pursued for the electrostatic modeling are predominantly 
analytical or numerical domain approaches such as the Finite Element Method
(FEM) or surface integral approaches such as the Boundary Element Method
(BEM) \cite{Senturia92}.
Among these, the first one, although very fast and accurate, can unfortunately
be implemented in a very limited range of two dimensional geometries. 
There have also been various attempts at providing
analytic expression for variation of capacitance for example, in 
parallel plate structure
as a function of the variation of geometrical properties in
\cite{Leus2004} and the references therein. The numerical approaches on the
other hand are capable of taking care of 3D geometry and thus can provide
more detailed and realistic estimates. The FEM approach is particularly
very flexible in terms of 3D modeling, but demands large computational 
expenditure for achieving the required
accuracy. As a result, the BEM approach has turned out to be the more popular
since it produces results of good accuracy with relatively less
computational cost. The method has its own drawbacks such as loss of accuracy
in the near-field, necessity of special treatment to handle numerical and
physical singularities. 
For example, different special formulations for
thick, moderately thin and very thin plate have been devised to compute the
necessary properties such as surface charge density and capacitance 
of a parallel plate structure in order to handle the drawbacks of 
BEM \cite{Bao2004}. 

Several parametrized formulations have been devised with the help of the
numerical methods for fast computation of capacitance in MEMS structures
\cite{Batra2006,Meijs,Palmer} in order to evade the time consuming 
modeling,
computation and other numerical complexities associated with the FEM and BEM.
However, these formulations are found to be restricted to specific parameter
space as well as 2D geometry.

All the drawbacks associated with FEM and BEM have been
removed to a large extent in a novel approach of BEM formulation 
devised by 
ourselves using the analytic solution of Green's function form of 
the potential due to a uniform charge distribution on a boundary element 
\cite{EABE2006,EMTM2N07b}. The solver based on this nearly exact BEM
(neBEM) formulations has been found to excel in electrostatic analysis of 
MEMS yielding very precise results throughout the domain of interest 
and at a very close proximity to any surface which are well known handicaps
of conventional BEM.
In \cite{EABE2006},
the effect of geometric parameters like thickness of each plate and the 
gap between them on the charge
density distribution and the capacitance of a parallel plate structure 
were studied. 
Here, a similar study has been carried out for a narrow beam structure
which has already
been identified to be strongly influenced by fringe field. 
For example, very strong fringe
field effect has been observed in vertical comb-drive actuator in comparison
to parallel plate actuator \cite{Hah2002} stressing the
need for carrying out such a study. Besides, it has been reported in \cite{Ma}
that even for a plate configuration, when the size of the electrod is not much
bigger than the gap, the fringe effect can not be ignored any more. This, of
course, was reflected in our earlier studies as well.
In addition to studying the effect of the various
geometric parameters such as the length, width, height and
gap between the two beams on the charge density distribution and capacitance,
the accuracy of the various approximate expressions available
for fast estimation of capacitance of comb-like structure relevant to MEMS
devices has also been checked out. The reasons of failure of such expressions 
have been discussed and possible ways of improvement have been suggested.

\section{\large{\textbf{Validation of neBEM}}}
\begin{figure}[hbt] 
\begin{center}
\includegraphics[height=2in,width=3in]{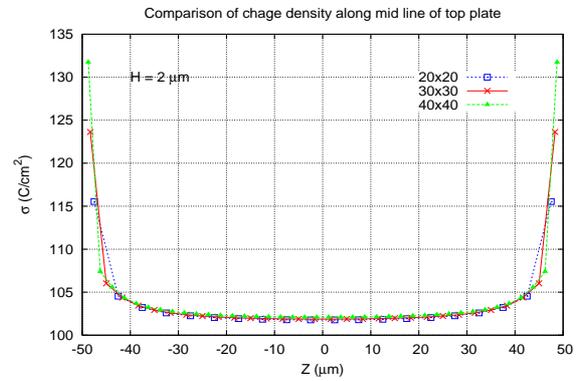}
\caption{\label{fig:ChargeDen} Surface charge density distribution on the 
upper plate of parallel plate configuration with gap $2\mu m$}
\end{center}
\end{figure}

The results of electrostatic force and capacitance for a parallel 
plate capacitor
provided by neBEM have been compared to that obtained with 
Coventor's ARCHITECT
software and its Parametrized ElectroMechanical model (PEM) \cite{Coventor}.
The parallel plate configuration consists of a square plate of side, 
$L_p=100\mu m$ and
thickness, $t_p=2\mu m$ separated by a distance $H$ from a square 
electrode with side,
$L_e=100\mu m$, thickness, $t_e=
0.45\mu m$. The analytical solution of the electrostatic force can be 
written as follows.
\begin{equation}
F_y = {V^2 \over 2} {\partial C \over {\partial H}} = -{V^2 \over 2} 
{\epsilon L_p^2 \over {H^2}}
\end{equation}
where $\epsilon = \epsilon_0 \epsilon_r$; 
the permittivity constant is $\epsilon_0=8.85\times10^{-12} F/m$ and $\epsilon_r=1$ 
is the relative permittivity of the medium.
\begin{figure}[hbt] 
\begin{center}
\includegraphics[height=2in,width=3in]{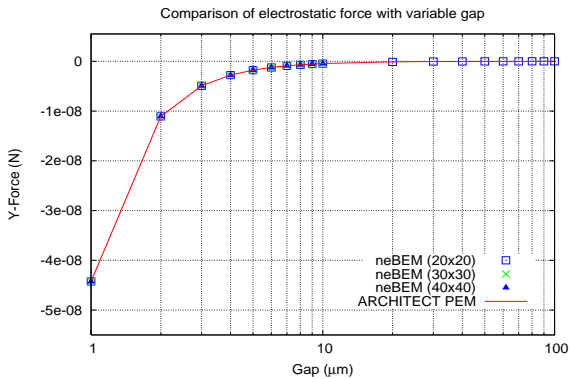}
\caption{\label{fig:CovForceComp} Electrostatic force acting on upper plate in
a parallel plate configuration}
\end{center}
\end{figure}
The ARCHITECT PEM calculation has shown a nice agreement with the 
analytical solution as seen from fig.\ref{fig:CovForceComp}. It should be 
mentioned here that the numerical convergence of the solution while 
using the BEM needs close inspection. It has
been experienced that the 
higher mesh refinement in BEM which produces convergent
capacitance values may not necessarily yield convergent force 
\cite{Iyer02}. In case
of neBEM, the numerical convergence of the results has been tested with 
mesh refinement as well as monitoring the associated charge distributions.
It has been found that even with relatively coarse a 
discretization scheme of $800$ elements only, the convergent force 
values could be achieved which hardly improved
by refining the scheme. The results produced with different discretization 
schemes have been depicted in 
fig.\ref{fig:CovForceComp}.
The charge density distribution on the top plate for a gap of $2\mu m$ has been shown in
fig.\ref{fig:ChargeDen} as a typical case.

\begin{figure}[hbt] 
\begin{center}
\includegraphics[height=2in,width=3in]{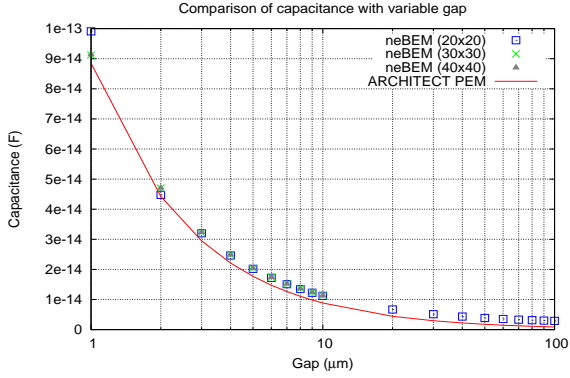}
\caption{\label{fig:CovCapComp} Capacitance calculated for a 
parallel plate configuration}
\end{center}
\end{figure}

The capacitance values calculated following the neBEM has shown a 
difference with what
predicted by the PEM calculation as evident from fig.\ref{fig:CovCapComp}. 
It is because the PEM has calculated the capacitance
following the analytical expression of parallel plate capacitance as follows
\begin{equation}
C = {\epsilon A \over H} = {\epsilon L_p^2 \over H}
\end{equation}
which neglect any fringe field contribution while the neBEM has well 
accounted for that effect in its calculation.

\section{\large{\textbf{Geometry of Narrow beam structures}}}
\begin{figure}[hbt] 
\begin{center}
\includegraphics[height=2in,width=3in]{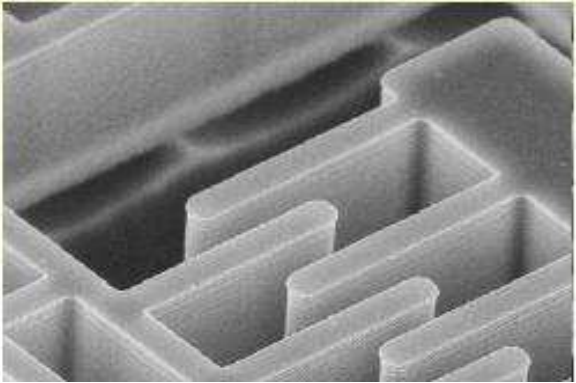}
\caption{\label{fig:CombDrive} Comb drives used as position
sensor (Photo courtsey:Kionix Inc.)}
\end{center}
\end{figure}

\begin{figure}[hbt] 
\begin{center}
\includegraphics[height=3in,width=3in]{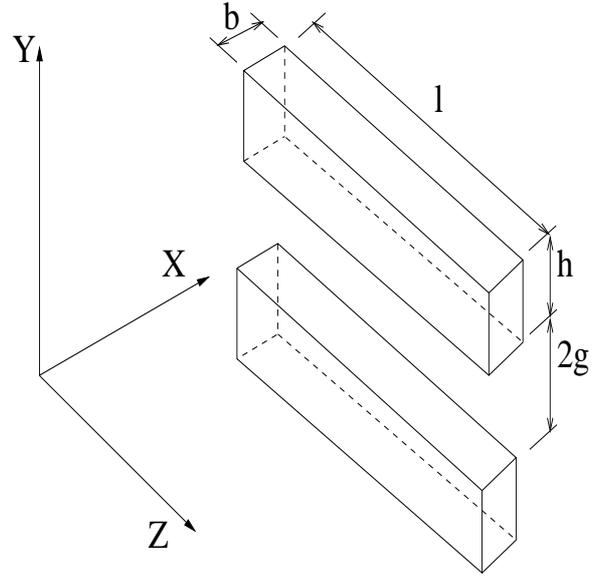}
\caption{\label{fig:GeomNarrowBeam} Narrow beam geometry considered for the
present calculations}
\end{center}
\end{figure}

\noindent{Narrow} beam structures of wide geometric variations have been 
used in MEMS
devices. As an example, in fig.\ref{fig:CombDrive} presented is a comb drive
which is actually used as a position sensor in MEMS systems.
We have considered a much simplified geometry (Fig.\ref{fig:GeomNarrowBeam})
which,
nevertheless, retains the basic characteristics of such structures. 
Here, the length, breadth, height of the beam are denoted by $l$, $b$
and $h$ respectively, and the half-gap between the two beams by $g$.
It can be seen from
various references in the published literature that $l$ can range from
$mm$ to tens of $\mu m$. The breadth, $b$ and height, $h$ can be
as small as $2\mu m$ and
$4\mu m$. The gap, $g$ also has a wide range of variation from tens of $\mu m$
to just $1\mu m$. There can be devices where these wide variations are even
more extended. Instead of trying to cover the whole range of the parameters, we
consider narrow beams of $l=150\mu m$, $b$ varying from $100\mu m$ 
to $2\mu m$, height $h=10\mu m$ and $g$ varying from $20\mu m$
to $1\mu m$. 

As is obvious from the above range of dimensions, effect of three 
dimensionality
in the narrow beam structures is likely to be even more significant than the
parallel plate structures. In many cases of practical interest, even approaches
based on reducing the 3D problem to a
2D one, and then recovering the 3D solution as developed in 
\cite{Suresh2006} is
unlikely to produce good results, especially when, according to some 
researchers
an accuracy of the order of $99\%$ is what should be acceptable 
in estimating the
different parameters related to MEMS devices. Moreover, approaches based on
reducing dimensions as above are likely to face problems when the cross-section
of the structure varies in the direction of sweep. Thus, these efficient
approaches can not be counted on for solving problems of a very general nature.
As a result, from the very outset, we prepared ourselves to handle the problem
in complete 3D. According to
us, rather than evading the 3D nature of the problem, more
important is to develop a fast and precise solver. The question of possible
parametrization in various ranges, if possible, can of course help in reducing
computational expenses and also help in saving time to a very great extent. But,
even to build proper parametric dependence, we need to solve the 3D problem to
an acceptable accuracy. 

\section{\large{\textbf{Results and discussions}}}
\noindent{To} study the variation of capacitance depending upon various
geometric parameters in a parallel narrow beam structure, a convention 
similar to \cite{Batra2006} has been adopted where,
in order to facilitate parametrization, the followings have been defined
\[
\beta = \frac{h}{b}, \,\,\, \eta = \frac{h}{g}
\]
In addition to the above, another dimensionless parameter
has been included.
\[
\lambda = \frac{l}{b}
\]
According to these parameters, the problem geometry varies from 
$0.1 \le \beta \le 5$,
$0.5 \le \eta \le 10$ and $1.5 \le \lambda \le 75$. The variation 
can be huge, but
a relatively narrow range has been considered to facilitate data
interpretation and analysis.

\subsection{\large{\textbf{Variation of capacitance}}}
\begin{figure}[hbt] 
\begin{center}
\includegraphics[height=2in,width=3in]{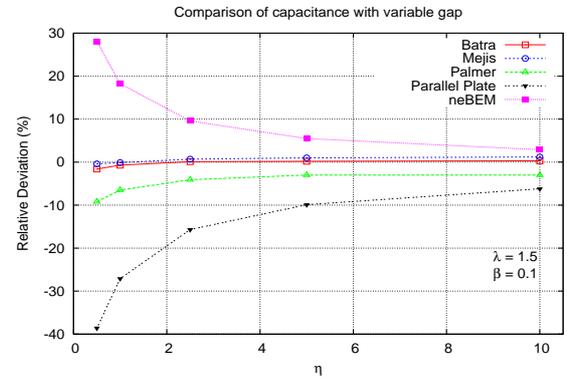}
\caption{\label{fig:CapCompGap} Relative deviation calculated capacitance
per unit length w.r.t MoM value}
\end{center}
\end{figure}

\noindent{To} study the variation of capacitance per unit length with different geometric parameters, the results
computed by the neBEM solver has been compared
with those calculated using various empirical parametric formulations presented in
\cite{Batra2006}. In Table \ref{table:CapComp}, the percent deviation incurred 
by various methods of estimation as well as the neBEM with 
respect to Method of Moment (MoM) calculation has been tabulated.
A column of $\lambda$ has been added in the
table to examine the effect of finite length of
the beam which has not been considered in the parametric calculations.
It may be recounted that for the present computations, $l$ and $h$ have been
kept fixed at $150\mu m$ and $10\mu m$. The parameter
$b$ has varied from $100\mu m$ to $2\mu m$, while $g$ from
$20\mu m$ to $1\mu m$ (Fig.\ref{fig:GeomNarrowBeam}.
\begin{table*}
\centering
\caption{\label{table:CapComp}Comparison of Capacitance per Unit length 
due to variation in geometric parameters}
\begin{tabular}{| c | c | c | c | c | c | c | c | c | c |}
\hline
$\lambda$ & $\beta$ & $\eta$ & MoM & Present & \cite{Batra2006} & \cite{Meijs}
& \cite{Palmer} & Par Plate & Present \\
\hline
1.5 & 0.1 & 0.5 &   8.14 &  10.43 & -1.6 & -0.4 & -9.2 & -38.6 & 28.13 \\
\hline
1.5 & 0.1 & 1.0 &  13.71 &  16.21 & -0.7 & <0.1 & -6.5 & -27.1 & 18.23 \\
\hline
1.5 & 0.1 & 2.5 &  29.61 &  32.45 & <0.1 &  0.7 & -4.1 & -15.7 &  9.59 \\
\hline
1.5 & 0.1 & 5.0 &  55.38 &  58.42 &  0.2 &  1.0 & -3.0 &  -9.9 &  5.49 \\
\hline
1.5 & 0.1 & 10. & 106.19 & 109.34 &  0.3 &  1.2 &  2.2 &  -6.2 &  2.91 \\
\hline
3.0 & 0.2 & 0.5 &   5.37 &   6.70 & -1.1 & -0.4 & -17.2& -53.5 & 24.76 \\
\hline
3.0 & 0.2 & 1.0 &   8.42 &   9.80 & -0.4 & -0.1 & -12.3& -40.7 & 16.39 \\
\hline
3.0 & 0.2 & 2.5 &  16.80 &  18.29 & <0.1 &  0.8 & -8.0 & -25.6 &  8.89 \\
\hline
3.0 & 0.2 & 5.0 &  30.06 &  31.59 &  0.3 &  1.5 & -5.6 & -16.9 &  5.09 \\
\hline
3.0 & 0.2 & 10. &  55.86 &  57.32 &  0.5 &  1.9 & -4.0 & -10.9 &  2.61 \\
\hline
7.5 & 0.5 & 0.5 &   3.61 &   4.34 & <0.1 & -0.8 & -34.5& -72.3 & 20.22 \\
\hline
7.5 & 0.5 & 1.0 &   5.12 &   5.82 &  0.4 & -0.6 & -25.7& -61.0 & 13.67 \\
\hline
7.5 & 0.5 & 2.5 &   8.97 &   9.69 &  0.4 & -0.7 & -17.7& -44.3 &  8.03 \\
\hline
7.5 & 0.5 & 5.0 &  14.71 &  15.41 &  0.6 &  2.1 & -12.9& -32.1 &  4.75 \\
\hline
7.5 & 0.5 & 10. &  25.50 &  26.09 &  0.9 &  3.4 & -9.1 & -21.9 &  2.31 \\
\hline
15. & 1.0 & 0.5 &   2.96 &   3.48 &  0.6 & -1.7 & -51.9& -83.1 & 17.57 \\
\hline
15. & 1.0 & 1.0 &   3.96 &   4.43 &  0.7 & -1.8 & -40.3& -74.8 & 11.87 \\
\hline
15. & 1.0 & 2.5 &   6.29 &   6.71 &  0.4 & -0.1 & -29.3& -60.2 &  6.68 \\
\hline
15. & 1.0 & 5.0 &   9.52 &   9.88 &  0.5 &  2.2 & -22.4& -47.5 &  3.78 \\
\hline
15. & 1.0 & 10. &  15.30 &  14.50 &  1.1 &  4.6 & -16.4& -34.9 & -5.23 \\
\hline
30. & 2.0 & 0.5 &   2.59 &   2.98 &  0.8 & -2.9 & -71.8& -90.4 & 15.06 \\
\hline
30. & 2.0 & 1.0 &   3.34 &   3.66 &  0.7 & -3.5 & -57.3& -85.0 &  9.58 \\
\hline
30. & 2.0 & 2.5 &   4.90 &   5.12 & <0.1 & -1.7 & -43.8& -74.5 &  4.49 \\
\hline
30. & 2.0 & 5.0 &   6.88 &   6.89 & <0.1 &  1.4 & -53.3& -63.7 &  1.45 \\
\hline
30. & 2.0 & 10. &  10.16 &   9.56 &  0.7 &  5.4 & -27.4& -50.9 & -5.90 \\
\hline
75. & 5.0 & 0.5 &   2.34 &   2.62 &  0.3 & -5.2 &  Inf & -95.7 & 11.96 \\
\hline
75. & 5.0 & 1.0 &   2.92 &   3.14 & -0.1 & -6.2 & -81.5& -93.2 &  7.01 \\
\hline
75. & 5.0 & 2.5 &   4.03 &   4.14 & -1.4 & -4.8 & -64.7& -87.6 &  2.73 \\
\hline
75. & 5.0 & 5.0 &   5.25 &   5.22 & -1.6 & -1.0 & -55.0& -81.0 &  0.57 \\
\hline
75. & 5.0 & 10. &   7.04 &   6.62 & -0.7 &  4.9 & -45.9& -71.6 & -5.96 \\
\hline
\end{tabular}
\end{table*}
The deviation of the calculations w.r.t MoM values has been plotted as a
function of $\eta$ (inverse gap) in fig.\ref{fig:CapCompGap}.
The wide variation between the neBEM and parametric results is, in fact, 
expected because the parametric formulation has been devised for 2D 
geometry whereas the neBEM takes care of 3D one. Whenever $\eta$ is small
(large gap), for small values of $\lambda$ (large $b$ for a fixed $l$),
the parametric assumption
breaks down resulting in large difference between the 2D and 3D results which
reflects the effect of fringing field. However, for the same $\lambda$, the
difference reduces considerably when the gap between the two structures is
reduced (larger $\eta$). Obviously, the disagreement between the 2D and 3D
results improves when $\lambda$ becomes larger (smaller $b$) for a fixed
$\eta$ reflecting the reduction in fringe field effect.
The MoM
result, and thus those due to \cite{Batra2006} are likely to be 
correct only if
the length is much larger than all other dimensions of the beam. This,
unfortunately, cannot be very realistic a picture for every narrow beam
structure used in MEMS. It is conceivable from the results that all
the lengths of the device play important role, and for a true measure of
capacitance and related properties, none of these can really be neglected
unless we stick to a reasonably narrow parameter window. The window considered
above, according to us, is not narrow enough.

\subsection{\large{\textbf{Variation of charge density distribution}}}
\begin{figure}[hbt] 
\begin{center}
\includegraphics[height=2in,width=3in]{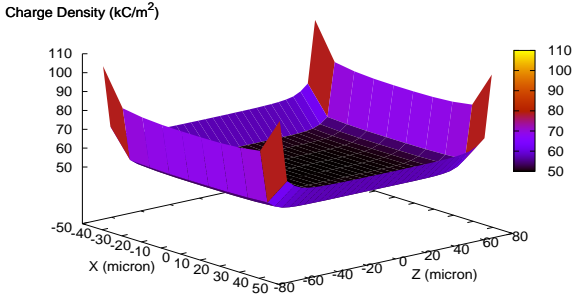}
\caption{\label{fig:ChDen} Surface charge density distribution on a surface 
facing the gap}
\end{center}
\end{figure}

\begin{figure}[hbt] 
\begin{center}
\includegraphics[height=2in,width=3in]{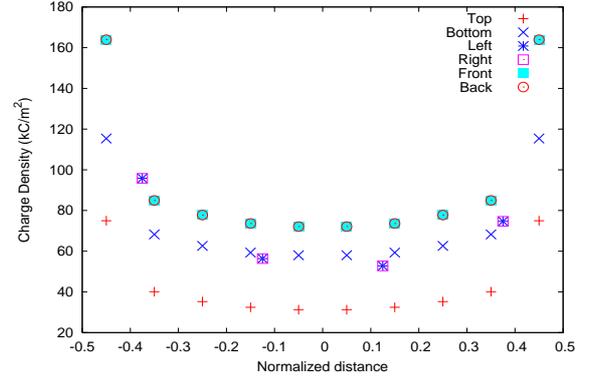}
\caption{\label{fig:ChDenCompLargeGap} Surface charge density distribution on
each surface of the upper beam with gap $20\mu m$}
\end{center}
\end{figure}

\begin{figure}[hbt] 
\begin{center}
\includegraphics[height=2in,width=3in]{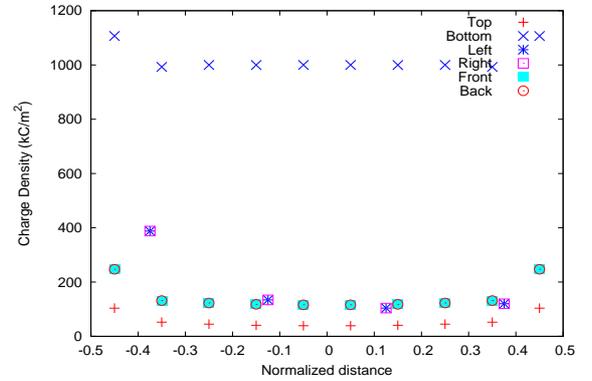}
\caption{\label{fig:ChDenCompSmallGap} Surface charge density distribution on
each surface of the upper beam with gap $1\mu m$}
\end{center}
\end{figure}

\noindent{There} is a large increase in charge density
near the edges and corners on each surface. Since the electrostatic force 
depends on this charge density distribution directly,
the finite nature of the length can cause error while predicting
these important properties of an MEMS device.
It may be safer to critically examine the parameter window of
interest and only then decide regarding the 2D or 3D nature of the problem. 
The surface charge density on the surfaces facing the gap and away from it
have been illustrated in
fig.\ref{fig:ChDen} and fig.\ref{fig:ChDenaway} which shows that the
charge density on surfaces not facing the gap is considerably smaller.
\begin{figure}[hbt] 
\begin{center}
\includegraphics[height=2in,width=3in]{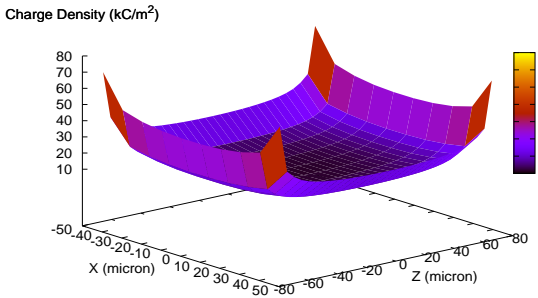}
\caption{\label{fig:ChDenAway} Surface charge density distribution on a surface
away from the gap}
\end{center}
\end{figure}
In order to emphasize the differences of the charge densities on the various
surfaces of the narrow beams, in figs.\ref{fig:ChDenCompLargeGap} and
\ref{fig:ChDenCompSmallGap}, the charge densities along the
mid-sections of each surface of a beam for the largest gap, $g=20\mu m$ and
the smallest gap, $g=1\mu m$ have been presented. 
The breadth of the beam has been considered to be
$b=20\mu m$ which is a representative one. Naturally, the charge density is
expected to vary as $b$ changes from $100\mu m$ to $2\mu m$. Please note that
in these figures, the charge densities on the upper beam of the
overall structure have been plotted. According to this
figure, the top surface is the one that is away from the gap, while the bottom
one faces the gap and the other beam structure. The left and right
surfaces are at constant values of X, while the front and back surfaces are at
constant values of Z. The variations of charge density for the top and bottom
surfaces are along X, left and right surfaces are along Y while for the front
and back surfaces are along the X axis. All the distances have been normalized
with respect to the length in the corresponding direction for the convenience
of both presentation and interpretation.

When the gap is large, the charge density distribution is found to be quite
even for many of the surfaces except the surface far away from the gap owing
to large fringe field effect. It
may be noted here that the symmetric nature of the problem is accurately
reflected in the presented results. Thus, the values obtained for left-right
and front-back surfaces are completely indistinguishable from each other. The
left and right surfaces have an asymmetric distribution of charge density with
respect to Y, which also is expected. Lesser values of Y in these cases implies
proximity to the gap and thus have larger values of charge density. Moreover,
the existence of the edges are clearly visible in all the
data points. It should be noticed here that all the surfaces having accumulated
quite even charges contribute significantly in the total charge content, thus
indicating the consideration of 3D nature of the problem.

The case where the gap between the beams is very small in comparison to all the
other dimensions, the
average charge density of the surface facing the gap is much larger than the
surface charge density of the all the surfaces of the beam implying less fringe
field effect. In addition, since
the area of this surface is large in comparison to most of the other surfaces,
its contribution towards the total charge content of the beam completely
dominates over the contribution of the other surfaces. This indicates that
at this situation, various electrostatic properties can be approximated by
two-dimensional configurations to a large extent. 

From the previous discussions, it has seemed to us that the very approach 
of supplying an expression that will
work throughout a large parameter space is bound to face difficulties. It may
be far more prudent to divide the parameter space in smaller sub-regions and
try to find appropriate relations valid in that sub-space of parameters. In
terms of computation, this is not likely to be burdensome as our 
experience with
the neBEM solver suggests. Moreover, the expressions are likely to be
far more precise in predicting the capacitance of MEMS structures. However, 
the very task of finding these expressions may be lengthy and laborious.
We plan to carry
out a thorough study in this direction in the near future.

\section{\textbf{\large{Conclusions}}}
\noindent{A} 3D computation of capacitance of a MEMS parallel 
narrow beam structure has been presented in this work. 
The results have been compared with several existing
estimates which are essentially 2D in nature. Significant variation in the
results have been observed and the reasons behind these 
discrepancies have been
discussed by analyzing the charge densities on the various surfaces of a beam
with the variation of the gap in the device. The effect of finite length, even
for narrow beams, has been found to be quite large in many areas of the
parameter space covered. It has been finally concluded that it may be 
difficult to
find a single expression that will represent the variation of capacitance with
geometrical parameters covering a large parameter space.

\vspace{24pt}
\noindent{\textbf{\large{Acknowledgements}}}\\
We would like to thank Professor Bikas Sinha, Director, SINP and Professor
Sudeb Bhattacharya, Head, INO Section, SINP for their support 
and encouragement during the course of this work.


\begin{thebibliography}{20}

\bibitem {Senturia92}
Senturia, S.D., Harris, R.M., Johnson, B.P., Kim, S., Nabors, K., Shulman, M.A. and White, J.K.,
1992,
"A computer-aided design system for microelectromechanical systems (MEMCAD)",
\textit{J Micro Electro Mech Syst}, \textbf{1}, pp.3-13.

\bibitem{Leus2004}
Leus, V. and Elata, D.,
2004,
"Fringing field effect in electrostatic actuators",
\textit{Technical report ETR-2004-2}, Technion - Israel Institute of 
Technology, Faculty of Mechanical Engineering, Israel, 15 pages.

\bibitem{Bao2004}
Bao, Z. and Mukherjee, S.,
2004,
"Electrostatic BEM for MEMS with thin conducting plates and shells",
\textit{Engg Analysis Bound Elem}, \textbf{28}, pp.1427-1435.

\bibitem{Batra2006}
Batra,R.C., Porfiri,M. and Spinello,D.,
2006,
"Electromechanical model of electrically actuated narrow microbeams",
\textit{J Microelectromechanical Systems}, Accepted for publication in 2006.

\bibitem{Meijs}
Meijs,N.V.D. and Fokkema,J.T.,
1984,
"VLSI circuit reconstruction from mask topology",
\textit{Integration}, \textbf{2}, pp.85-119.

\bibitem{Palmer}
Palmer,H.B.,
1937,
"Capacitance of a parallel plate capacitor by the Schwartz-Christoffel
transformation",
\textit{Trans Amer Inst Elect Eng}, \textbf{56}, pp.363-366.

\bibitem{EABE2006}
Mukhopadhyay, S. and Majumdar, N.,
2006,
"Computation of 3D MEMS electrostatics using a nearly exact BEM solver",
\textit{Engg Analysis Boun Elem}, \textbf{30}, pp.687-696.

\bibitem{EMTM2N07b}
Mukhopadhyay, S. and Majumdar, N.,
2007,
"Use of rectangular and triangular elements for nearly exact BEM solutions",
\textit{Proc. of Intl. Conf. on Emerging Mechanical Technology-
Macro to Nano (EMTM2N-2007)} (ISBN 81-904262-8-1), BITS-PILANI, India, 
February 16-18, pp.107-114.

\bibitem{Hah2002}
Hah, D., Huang, S., Nguyen, H., Chang, H., Tsai, J-C., Wu, M.C. and Toshiyoshi, H.,
2006,
"Low voltage MEMS analog micromirror arrays with hidden vertical comb-drive
actuators",
\textit{Solid-State Sensor, Actuator and Microsystems Workshop},
Hilton Head Island, South Carolina, June 2-6, pp.11-14.

\bibitem{Ma}
Ma, Y.,
2004,
"Optimal MEMS plate design and control for large channel count optical
switches",
\textit{Ph.D. Thesis, Faculty of Graduate School of the University of 
Maryland}, College Park, Maryland, USA .

\bibitem{Coventor}
\textit{http://www.Coventor.com/media/fem\_comparisons/parallel\_plate.pdf}

\bibitem{Iyer02}
Iyer, S., Lakdawala H., Mukherjee T. and Fedder G. K.,
2002,
"Modeling methodology for a CMOS-MEMS electrostatic comb"
\textit{Proceedings of SPIE: Design, Test, Integration and Packaging of
MEMS/MOEMS 2002}, \textbf{4755}, pp.114-125.

\bibitem{Suresh2006}
Sirpotdar, A. and Suresh, K.,
2006,
"A 2D model that accounts for 3D fringing in MEMS devices",
\textit{Journal of Engg Analysis Boun Elem}, accepted January 2006.




\end{thebibliography}
\end{document}